\documentclass[a4paper,1pt]{article}

\usepackage{amsmath, amssymb, amsthm, amsfonts, cases}
\usepackage{mathrsfs}
\usepackage{url}
\usepackage{authblk}
\usepackage[usenames]{color}
\usepackage{float}
\usepackage{subfigure}
\usepackage{caption2}
\usepackage{geometry}
\usepackage{multirow}
\usepackage{graphics}
\usepackage{epsfig}

\usepackage{epstopdf}
\allowdisplaybreaks

\newtheorem{theorem}{Theorem}[section]

\newtheorem{proposition}{Proposition}[section]

\newtheorem{remark}{Remark}[section]

\makeatletter\@addtoreset{equation}{section} \makeatother

\begin{document}
\title{Optimal investment problem with M-CEV model: closed form solution and applications to the algorithmic trading.}

\author{Dmitry Muravey\footnote{e-mail:d.muravey@mail.ru. This work was sponsored by Russian Science Foundation, project number 15-11-30042.}}
\affil{Department of Probability, Steklov Mathematical Institute RAS, Moscow, Russia}

\date{}
\maketitle

\begin{abstract}
\noindent
This paper studies an optimal investment problem under M-CEV with power utility function. Using Laplace transform we obtain an explicit expression for the optimal strategy in terms of confluent hypergeometric functions. For the  representations obtained, we derive asymptotic and approximation formulas containing only elementary functions and continued fractions. These formulas allow us to analyze the impact of the model's parameters and the effects of their misspecification. In addition we propose extensions to our results that are applicable to algorithmic trading.
\end{abstract}

\section{Introduction}
\subsection{Motivation}
Many academic papers about optimal investment problems assume that the asset price follows geometric Brownian Motion(GBM). However, there are a lot of empirical studies showing this simple model does not properly fit to real market data. Known drawbacks are the following: GBM model does not capture volatility smile/skew effects; they ignore the probability of the underlying's default; the constant coefficients do not allow calibration of this model to the real term structure of interest rates and dividend yields etc. Our motivation is to extend the results of GBM models to a more realistic model. In order to obtain a more realistic fit to the market data we can use more sophisticated models based, for example, on Levy processes or on fractional Brownian motion. But although their dynamics are more realistic, these complicated models are not usually analytically tractable. Hence quantitative analysis is complicated and any qualitative analysis is impossible. We must try compromise between realistic modelling and the availability of analytical or quasi-analytical expressions. 
\par
In this paper we solve an optimization problem assuming the Modified Constant Elasticity of Variance (i.e. M-CEV) model for the asset's price and a power utility over the final wealth for a finite horizon agent. This model was introduced in Heath and Platen (2002) and is a natural extension of the famous CEV model (see Cox(1975)). We choose this model for the following reasons: this model captures the volatility smile effect; allows non-zero probability of the underlying's default (M-CEV process can touch zero while GBM is always positive); and it is analytically tractable. Also this model is applicable to algorithmic trading strategies because the M-CEV process has a mean-reversion property for some of the model's parameters. Let us mention that the time-dependent extension of this model can be found in Linetsky and Carr (2006). For the M-CEV model we obtain a closed-form solution in terms of confluent hypergeometric functions. Despite the availability of many numerical solvers (i.e. PDE solvers or Monte-Carlo) explicit formulas are still relevant. There are several reasons to pursue a closed form solution: first, they show dependencies between model parameters and optimal policy, therefore we can obtain some non-trivial qualitative effects. Second, properly programmed closed-form solutions give faster and more efficient code than a lot of available numerical solvers(PDE solvers or Monte-Carlo). In addition, simple tractable models can serve as a benchmark in practical situations. Quite often, practitioners prefer to introduce {\it ad hoc} corrections to a simple model than to use a more involved model with a large number of parameters.
\par
Another important point is the utility choice. There are some popular utility functions considered in the literature: logarithmic, power and exponential. Obviously, each utility gives a different optimal strategy that maximizes expected utility over the terminal wealth. It is well known that the optimal strategy in the case of a logarithmic utility does not depend on the time to the end of the investing period and the trading rules of an exponential utility investor is not sensitive to the current wealth (see Merton (1990)). In order to capture time and wealth dependencies we choose a power utility.

\subsection{Previous research}
\par
There are a lot of papers about similar problems: T. Zariphopoulou (2001) considered the problem for stochastic volatility models and derived optimal policy as the solution of the parabolic PDE. Some closed-form solutions and asymptotic expansions for various models can be found in Kraft (2004), Chacko and Viceira (2005), Boguslavskaya and Muravey (2015). A detailed review of papers about closed-form solutions and asymptotics can be found in Chan and Sircar (2015).
\par
Applications of the utility maximization problems to algorithmic trading were discussed in Boguslavsky and Boguslavskaya (2004), Liu and Longstaff (2000). 
\subsection{The main results and structure of the paper}
\par
The main result of this paper is the closed form solution for the expected utility maximization in the finite horizon with power utility and M-CEV model. We derive asymptotic and approximation formulas containing only elementary functions and continued fractions. The structure of this paper is as follows: first we define the problem. Then we present a closed form solution for the M-CEV model. This is followed by the algorithm of numeric implementation and an analysis of parameter misspecification. Applications of the obtained results to algorithmic trading strategies are then given. All proofs are in Appendix \ref{sec:appendix}.

\section{Problem definition}\label{sec:problemdefinition}
\subsection{Model setup}
\par 
Consider a simple market consisting of a risk-free bond $B_t$ and a risky asset (i.e. stock) $S_t$. The bond and stock prices are driven by SDE: 
\begin{eqnarray}
\label{eq:Sdef}
dB_s &=& r(s) B_s ds, \quad\quad B_t = B >0, \nonumber \\
d S_s / S_s &=& [r(s) - q(s) + \lambda(S_s,s)] ds + \sigma(S_s,s) dW_s, \quad\quad S_t = S >0,
\end{eqnarray}
where $W_s$ is a standard Wiener process, $r(s) \geq 0 $, $q(s) \geq 0 $, $\sigma(S,s) > 0$  and $\lambda(S,t) \geq 0$ are the time-dependent risk-free interest rate, the time-dependent dividend yield, the time- and state- dependent instantaneous stock volatility, and the time- and state- dependent default intensity, respectively.The M-CEV model has the following specifications:
\begin{eqnarray}
\sigma(S_s,s) = aS^\beta, \quad \lambda(S,s) = b + c\sigma^2(S,s) = b + ca^2 S^{2\beta},
\quad q(s) = q, \quad r(s) = r,  \quad \alpha = r - q + b,
\end{eqnarray}
and defined by this corresponded SDE
\begin{eqnarray}
\label{eq:MCEV}
d S_s / S_s = \left[\alpha + ca^2 S^{2\beta} \right] ds + aS^{\beta} dW_s.
\end{eqnarray}
Let us mention that Heath and Platen considered model (\ref{eq:MCEV}) with $c = 1$. The case of $c \neq 1$ is not extension of original M-CEV model because this case can be reduced to the original model by a simple change of measure. We will use specification (\ref{eq:MCEV}) with $c \neq 1 $ to analyze the impact of parameter $c$ directly. The optimal investment problem can be treated in the general portfolio optimization framework. Assuming no market frictions and an absence of transaction costs, the wealth dynamics for a control $\pi_s$ is given by
\begin{eqnarray}
\label{eq:Xdef}
dX_s = r (X_s - \pi S_s) ds + \pi_s dS_s.
\end{eqnarray}
Here $\pi_s$ is the investor position in stock(i.e. the number of units of the asset held). We assume that there are no restrictions on $\pi_s$, so short selling is allowed and there are no marginal requirements on wealth $X_s$. We solve the expected terminal utility maximization problem for an agent with a prespecified time horizon $T$ and initial wealth $X_0 > 0$. The value function $J(X,S,t)$ is the expectation of the terminal utility conditional on the information available at time $t$ ($S_t =S$, $X_t=X$).

\begin{eqnarray}
\label{eq:Jdef}
   J(X, S, t) = \sup_{\pi} \mathbb{E} \, [U( X_T) \,\, | \,\, X_t = X,\, S_t = S],
\end{eqnarray}
where $U(X)$ is the power utility function
\begin{eqnarray}
\label{eq:utility}
U(x) = \frac{x^{\gamma}}{\gamma}.
\end{eqnarray}

\subsection{Known results}
In this section we provide some known results used later in this paper. The first result is about a reduction of the original problem (\ref{eq:Jdef}) with power utility (\ref{eq:utility}) to the Parabolic partial differential equation (PDE).
\begin{theorem}[Zariphopoulou]
Assume that the asset price process $S_t$ follows SDE
\begin{equation}
\label{zar process}
\left\{ {\begin{array}{l}
        dS_s / S_s =  \mu(V_s,s) ds + \sigma(V_s,s) dW_s^1, \quad S_t = S, \\
        dV_s = b(V_s,s)ds + a(V_s,t) d W_s^2, \quad V_t = v,
 \end{array}} \right.
\end{equation}
where $W_s^1$ and $W_s^2$ are correlated Wiener processes with coefficient $\rho$ and the investor has power utility function (\ref{eq:utility}). In these assumptions the value function (\ref{eq:Jdef}) can be represented as(i.e. distortion transformation)
\begin{eqnarray}
\label{eq:Jdistortion}
            J(X,S,v,t) = \frac{X^\gamma}{\gamma} f^{1/\delta} (v,t), \quad \quad \delta = 1+\rho^2\frac{\gamma }{1-\gamma}.
\end{eqnarray}
Function $f$ is a solution of the linear parabolic PDE boundary problem
\begin{equation}
\label{eq:fPDEZariphopoulouL}
\left\{ {\begin{array}{l}
f_t + \frac{1}{2} a^2(v,t) f_{vv} +
\left[b(v,t) +\rho\frac{\gamma (\mu(v,t)-r(t)) a(v,t)}{(1-\gamma)\sigma(v,t)}\right] f_v +
\frac{ \gamma \delta }{1-\gamma} \left[\frac{(\mu(v,t)-r(t))^2}{2\sigma^2(v,t)} + (1-\gamma)r \right]f =0, \\
f(v,T) = 1, 
	\end{array}} \right.
\end{equation}
Optimal policy $\pi^*(S_t,X_t,v_t,t)$ is given in the feedback form
\begin{eqnarray}
            \label{eq:piStarZariphopoulou}
            \pi^{*} (X, S, v, t) = \frac{X}{S(1-\gamma)} \left( \frac{\mu(v,t)-r(t)}{\sigma^2(v,t)} + \frac{\rho} {\delta} \frac{a(v,t) f_v (v,t)}{\sigma(v,t) f(v,t)} \right).
\end{eqnarray}
\end{theorem}
It is easy to show that the T.Zariphopoulou result can be applied to the M-CEV model (\ref{eq:MCEV}) by substitution
\begin{eqnarray}
\label{eq:substitutions2MCEV}
S=v, \quad \rho=1, \quad a(S,s) = S \sigma (S,s), \quad b(S,s) = S \mu(S,s).
\end{eqnarray}
\begin{proposition}
For M-CEV model the value function $J(X,S,t)$ is given by
\begin{eqnarray}
\label{eq:JdistortionMCEV}
J(X,S,t) = \frac{X^\gamma}{\gamma} f^{1/\delta}(S,t), \quad \delta = \frac{1}{1-\gamma}.
\end{eqnarray}
Function $f$ solves Cauchy problem 
\begin{equation}
\label{eq:MCEV_operator}
\left\{ {\begin{array}{l}
	\mathcal{L}f \equiv f_t + \frac{a^2 S^{2 \beta + 2 }}{2}  f_{vv} +
	\delta S[\alpha-\gamma r + c a^2 S^{2\beta} ] f_v +
\frac{\delta (1- \delta)}{2a^2} \left[\left(\alpha-r\right) S^{-\beta} + cS^{\beta} \right]^2 f + r\gamma \delta f =0, \\
	f(v,T) = 1, 
	\end{array}} \right.
\end{equation}
and optimal policy $\pi^*(X, S, t)$ is given by
\begin{eqnarray}
            \label{eq:piStarMCEV}
            \pi^{*} (X, S, t) = X \left( \delta \frac{\alpha-r + c a^2 S^{2\beta}}{a^2 S^{2\beta+1}} + \frac{f_S (v,t)}{f(v,t)} \right).
\end{eqnarray}
\end{proposition}
The main difficulty is to solve boundary problem (\ref{eq:MCEV_operator}).  In the next section we present a closed-form solution of (\ref{eq:MCEV_operator}) in terms of confluent hypergeometric functions.
\section{Main results}\label{sec:mainresults}
Consider the Cauchy problem  (\ref{eq:MCEV_operator}) with arbitrary initial function $f(S,T) = g(S)$. It is known that its solution can be represented as a convolution product with Green function $f_{G} (S, t; \xi)$
\begin{eqnarray}
\label{eq:Green_function}
f(S,t) = \int_{0}^{\infty} f_{G} (S, t; \xi) g(\xi) d\xi.
\end{eqnarray}
Using the Laplace transform method we obtain the explicit representation for Green function $f_G(S, t; \xi)$ in terms of Modified Bessel function $I_{\nu} (z)$ (for definition see Abramovitz and Stegun (1971)). Hence the solution of problem (\ref{eq:MCEV_operator}) can be easily obtained by application of formula (\ref{eq:Green_function})  with initial function $g(S) \equiv 1$. 
For convenience we will use scaled space and inverse time variables $z$ and $\tau$:
\begin{eqnarray}
\label{eq:zandtaudef}
z = \frac{\Lambda}{S^{2\beta}},
\quad \tau = a^2 \beta^2 \Lambda (T- t),
\quad \Lambda = \frac{\sqrt{\delta}}{a^2 |\beta|}
\sqrt{\alpha^2-\gamma r ^2 }, 
\end{eqnarray}
for function $f(S,t)$ we have the following representation
\begin{eqnarray}
\label{eq:Green_function_main}
f(S,t) = \int_{0}^{\infty} F_{G} (z, \tau; \xi) g\left( \left(\frac{\Lambda}{\xi} \right)^{1/2\beta} \right) d\xi.
\end{eqnarray}
In the next theorem we introduce explicit formulas for Green function $F_G(z, \tau; \xi)$. 
\begin{theorem} \label{thm:first} 
Green function $F_G(z, \tau; \xi)$ is given by
\begin{eqnarray}
\label{eq:Green_function_main_formula}
F_G(z,\tau;\xi) = \frac{1}{2} \exp\left\{ R\tau + Q(z-\xi) - \frac{(z+\xi)}{2}\coth(\tau)\right\} 
\left( \frac{z}{\xi}\right )^{\lambda + 1/2}  \frac{1}{\sinh(\tau)}I_{2\eta} \left(\frac{\sqrt{z\xi}}{\sinh(\tau)} \right),
\end{eqnarray}
where $\lambda$, $\eta$, $R$ and $Q$ are constants
\begin{eqnarray}
\label{eq:lambda&eta}
\lambda = -\frac{1}{2} - \frac{1}{2\beta} \left( \frac{1}{2} - \delta c \right), \quad\quad
\eta = \sqrt{\left(  \lambda + \frac{1}{2}\right)^2  + \frac{ \delta (1-\delta) c^2}{4 a^4 \beta^2}},
\end{eqnarray}
\begin{eqnarray}
\quad Q = \frac{\delta (\alpha - \gamma r)}{\Lambda \beta a^2}, \quad
R = \frac{r\delta}{a^2\beta^2 \Lambda} -2Q\lambda -\frac{\delta (1-\delta )(\alpha - r) c}{\Lambda a^4 \beta^2}.
\end{eqnarray}
\end{theorem}
Hence the solution of boundary problem (\ref{eq:MCEV_operator}) can be represented as   
\begin{eqnarray}
\label{eq:Green_function_main}
f(S,t) = \int_{0}^{\infty} F_{G} (z, \tau; \xi) d\xi.
\end{eqnarray}
We can perform these integrations explicitly by using the following relation (see Gradshteyn and Ryzhik (1980), formula 6.643.2) between Modified Bessel function $I_\nu(z)$ and Whittaker function $M_{\lambda, \eta}$(z) (see Abramowitz and Stegun (1973)) 
\begin{eqnarray}
\int_0^{\infty} x^{\mu -\frac{1}{2}}  e^{-\alpha x}  I_{2\nu} \left( 2 \beta \sqrt{x} \right) dx =
\frac{\Gamma \left( \mu + \nu  + \frac{1}{2}  \right)}{ \Gamma(2\nu+1)} \beta^{-1} e^{\frac{\beta^2}{2\alpha}} \alpha^{-\mu} M_{-\mu,\nu} \left( \frac{\beta^2}{\alpha }\right), \nonumber \\
{\rm Re \,}  \left( \mu + \nu +  \frac{1}{2} \right) > 0. \nonumber
\end{eqnarray}
In the result we have the following formula for function $f$
\begin{eqnarray}
\label{eq:exactfMCEV}
f(S,t) = {e^{R\tau + zB(\tau)} D^{\lambda} (\tau) }\frac{\Gamma(\eta -\lambda + 1/2)}{\Gamma(1+2\eta)} e^{-\frac{z}{2} A(\tau)}
\left( z A(\tau) \right)^\lambda M_{\lambda, \eta} \left( zA(\tau) \right),
\end{eqnarray}
where $\Gamma(x)$ is the Euler gamma function and functions $A(\tau)$, $B(\tau)$ and $D(\tau)$ are given by
\begin{eqnarray}
A(\tau) = \frac{1}{2 \sinh^2(\tau) [\coth(\tau) + Q]}, \quad
B(\tau) = \frac{Q^2 - 1}{2[\coth(\tau) + Q]}, \quad
D(\tau) = \sinh^2(\tau) [\coth(\tau) + Q]^2.
\end{eqnarray}
The expression for $ f_S /f $ is obtained by using differential rules for Whittaker functions (see Abramowitz and Stegun (1973))
\[
\left( z \frac{d}{dz} z \right)^n
\left( e^{-z/2} z^{k-1} M_{k,\mu} (z)\right) =
\frac{\Gamma(\mu + k + n + 1/2)}{\Gamma(\mu + k + 1/2)}
e^{-z/2} z^{k+n-1} M_{k+n,\mu} (z).
\]
Hence the optimal policy $\pi^*(X,S,t)$ is
\begin{eqnarray}
\label{eq:piStarExactMCEV}
\pi^*(X,S,t) =  X \left( \delta \frac{\alpha-r  + ca^2 S^{2\beta}}
{a^2 S^{2\beta + 1}}  + \left[ B(\tau) + \frac{\lambda + \eta + 1/2}{z}
 \frac{M_{\lambda+1, \eta}\left( A(\tau) z \right)}{M_{\lambda, \eta}\left(
 	 A(\tau) z \right)}\right] \frac{dz}{dS} \right) .
\end{eqnarray}
Using the following relation between Whittaker function and Kummer function  
\begin{eqnarray}
\label{eq:Whittaker2Kummer}
M_{\lambda, \eta} (x) = e^{-x/2} x^{1/2 + \eta}  \Psi(\theta, \omega, x), \quad 
\theta = 1/2 + \eta - \lambda, \quad \omega =1+2\eta
\end{eqnarray}
and compute derivative $dz/dS$ we obtain alternative formulas for $\pi^*(X,S,t)$: 
\begin{eqnarray}
\label{eq:piStarExactMCEVrearranged}
\pi^*(X,S,t) =  \frac{X}{S} \left[ \frac{ \delta (\alpha-r)/ a^2 - 2 \beta \Lambda B(\tau)}{S^{2\beta}}+ 
\delta c + 2\beta (\theta - \omega) \frac{\Psi\left(\theta - 1, \omega, \Lambda A(\tau)S^{-2\beta}\right)}
{ \Psi\left(\theta, \omega, \Lambda A(\tau)S^{-2\beta}\right)} \right] .
\end{eqnarray}

\section{Numerics} \label{sec:numerics}
\subsection{Numerical algorithm}
\par
\begin{figure}
	\begin{center}
		\resizebox*{18cm}{!}{\includegraphics{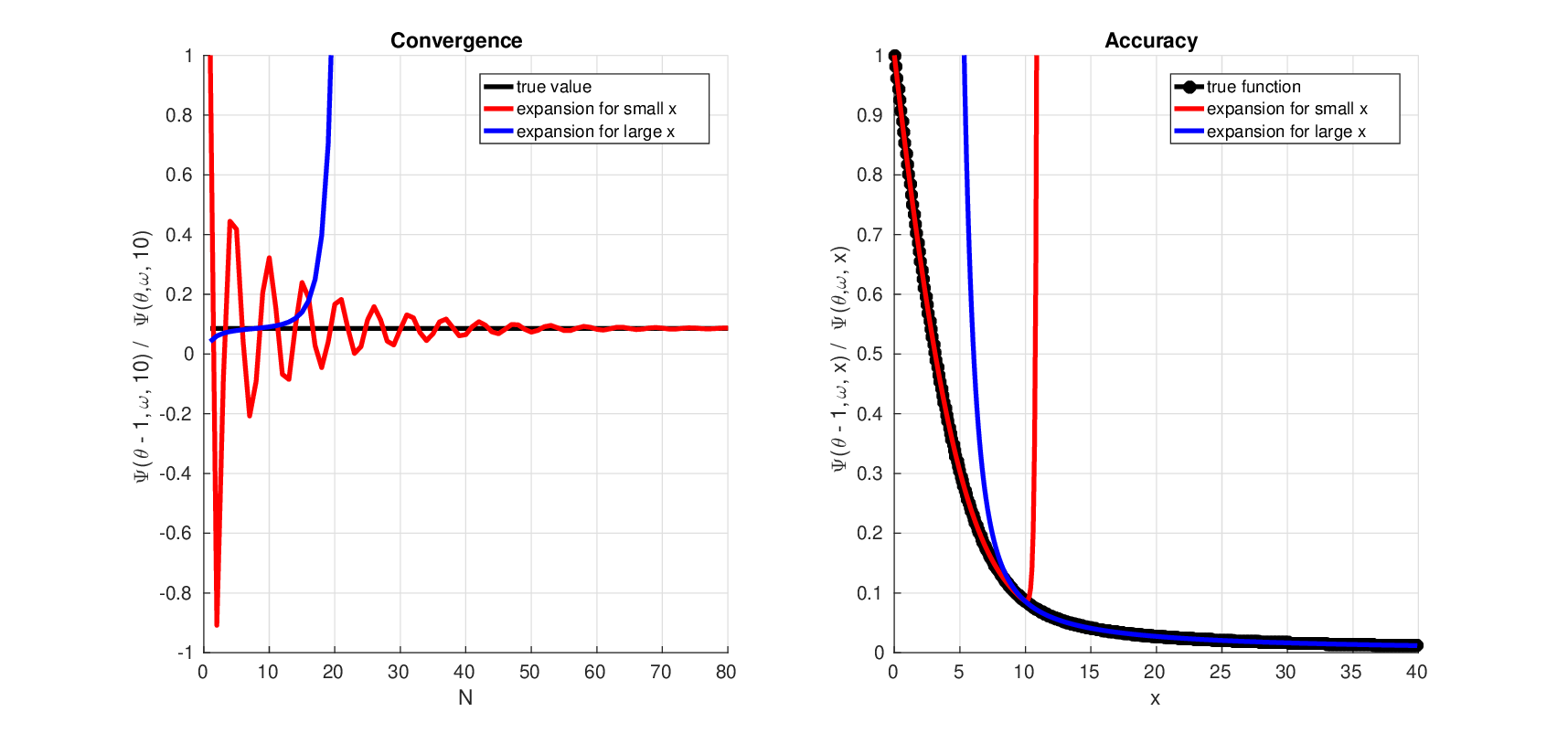}}
		\caption{\label{fig:3} Convergence rate (left sub-figure) and accuracy of approximations (right sub-figure).
			\label{fig:approx}}
	\end{center}
\end{figure}

 
If we want to build any quantitative trading strategy based on the obtained results we should have a numerical algorithm to compute expression \ref{eq:piStarExactMCEV} (or \ref{eq:piStarExactMCEVrearranged}) for any parameters. It consists of only elementary functions except the term 
\begin{eqnarray}
\label{eq:special_function_term}
\frac{\Psi(\theta - 1, \omega, x)}{\Psi(\theta, \omega, x)}, 
\quad \quad  \frac{M_{\lambda+1, \eta}\left( A(\tau) z \right)}{M_{\lambda, \eta}\left(
	A(\tau) z \right)}.
\end{eqnarray}
Obviously, computation of these special functions is not a problem for packages such as MATLAB or Mathematica. However, production codes are mostly written in C++ and we can not use these packages. In this context we must provide fast and efficient computation for this non-elementary term in a C++ environment. There are libraries containing numerical algorithms for special functions 
(e.g. C++ GSL package has numerics for the Kummer confluent hyper-geometric function used in 
(\ref{eq:special_function_term})). Hence we can compute 
(\ref{eq:special_function_term}) by the following scheme: if we have singularity in (\ref{eq:special_function_term}) we use asymptotic formulas(\ref{eq:asymptotic}), in other situations we use GSL. However, evaluations of Kummer functions can 
significantly slow down the computational speed of the algorithm and this approach is not suitable if speed is critical. In this section we provide a fast numerical scheme based on asymptotic expansions of term (\ref{eq:special_function_term}). The main idea is very simple. We construct two series expansions directly for term 
(\ref{eq:special_function_term}):                           
\begin{eqnarray}
\label{eq:series_exp_special_term}
\frac{\Psi(\theta - 1, \omega, x)}{\Psi(\theta, \omega, x)} = \sum_{s=0}^{\infty} c_s x^s = \frac{\theta - 1}{x}\sum_{s=0}^{\infty} d_s x^{-s}
\end{eqnarray}
and use first or second series depending on value of variable $x$. We compute approximation of series (\ref{eq:series_exp_special_term}) recursively with the following stopping criteria: we stop evaluations if the difference between $N+1$ and $N$ truncated series is sufficiently small. In the next theorem we provide explicit formulas for coefficients $c_s$ and $d_s$.      
\begin{theorem}
	\label{thm:second}
	Coefficients $c_s$ and $d_s$ in expansions (\ref{eq:series_exp_special_term}) are defined by recursive formulas 
\begin{eqnarray}
\label{eq:coef_c_and_d}
c_s = \frac{(\theta - 1)_s}{s!(\omega)_s} - \sum_{i+j = s}  \frac{c_j(\theta)_i}{i!(\omega)_i} , \quad
d_s = \frac{(2 -\theta)_s (\omega -\theta + 1)_s}{s!} - \sum_{i+j = s}  \frac{c_j(1 -\theta)_i (\omega -\theta)_i}{i!}.
\end{eqnarray}
\end{theorem}
Figure (\ref{fig:approx}) illustrates the convergence rate with fixed parameters $\theta$, $\omega$ and variable $x$ (left sub-figure) and accuracy of approximations (right sub-figure). For these tests we set $\theta = 5.24$ and $\omega = 1.42$. We illustrate convergence rate at the point $x = 10$ and for accuracy illustration we set $N = 80$ in expansions for small argument and set $N = 8$ for large.       
\par 
\begin{remark}
If we truncate series (\ref{eq:series_exp_special_term}) by 1 or 2 terms we obtain the following asymptotic for (\ref{eq:special_function_term})
\begin{eqnarray} \label{eq:asymptotic} 
\Psi(\theta-1,\omega, x) / \Psi(\theta,\omega, x)  \thicksim 1- x/\omega , \quad x\rightarrow 0, \nonumber \\
\Psi(\theta-1,\omega, x) / \Psi(\theta,\omega, x) \thicksim   (\theta-1)/x, \quad x \rightarrow \infty.
\end{eqnarray}
Moreover, recursive application of the following relation between Kummer functions (see Abramowitz and Stegun (1973))
\begin{eqnarray}
\label{eq:Kummer recursive}
(\omega - \theta)\Psi(\theta - 1, \omega, x) + 
(2\omega - \theta + x )\Psi(\theta, \omega, x) -
\theta \Psi(\theta + 1, \omega, x) = 0
\end{eqnarray}
turns out to the continued fraction representation
\begin{eqnarray} \label{eq:cont_fraction}
\Psi(\theta-1,\omega, x) / \Psi(\theta,\omega, x) = b_0 +
\cfrac{a_1}
{
	b_1 +
	\cfrac{a_2}
	{
		b_2 +...
	}
}
\end{eqnarray}
where
\begin{eqnarray}
a_n = \omega + n; \quad\quad b_n(x) = \frac{2\theta + 2n -\omega +  x}{\theta + n - \omega }.
\end{eqnarray}
\end{remark}

\subsection{Computational speed test}
\par
\begin{figure}
	\begin{center}
		\resizebox*{18cm}{!}{\includegraphics{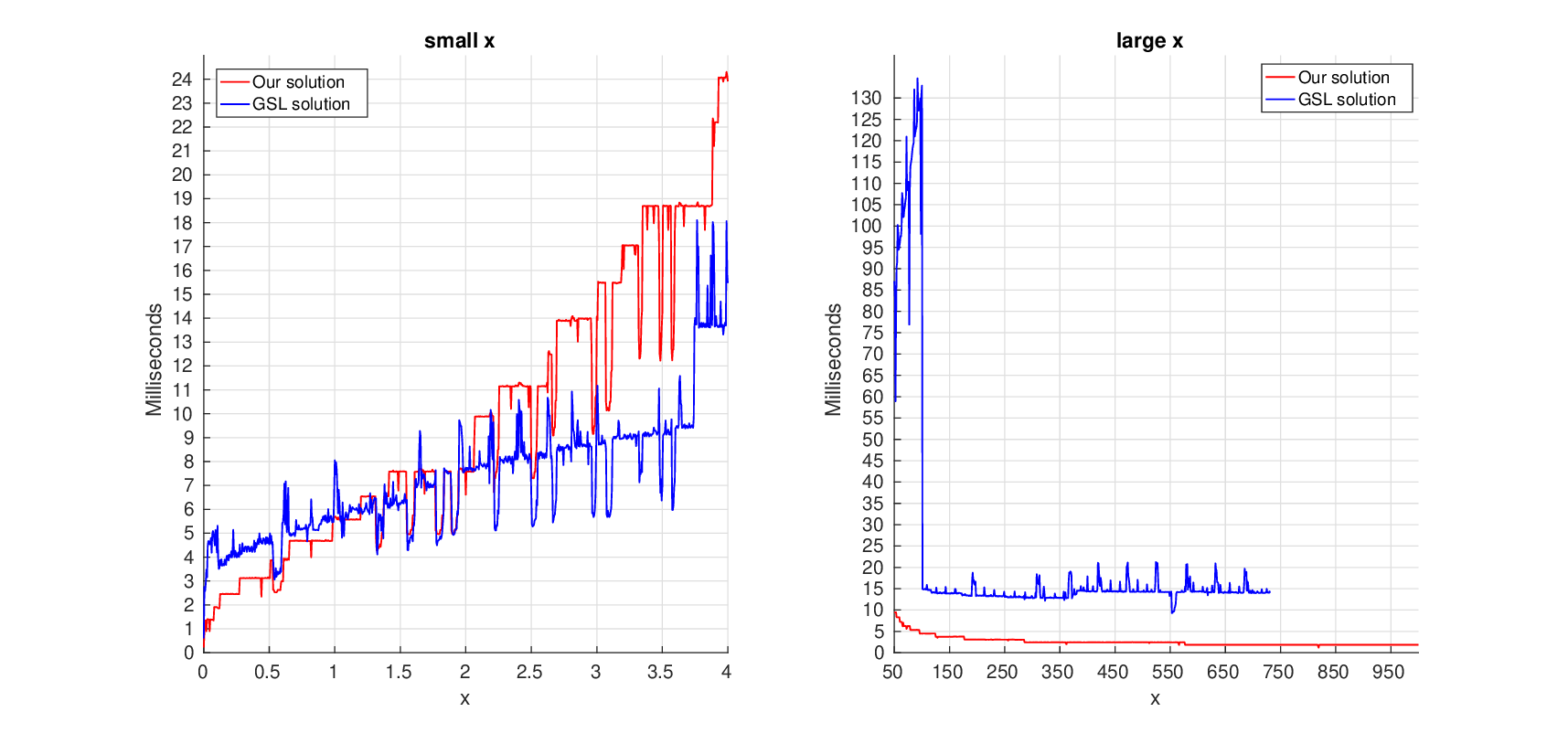}}
		\caption{\label{fig4} Computation speed test. Left and right sub-figures illustrate computational speeds for small and large argument.  
			\label{fig: speed_test}}
	\end{center}
\end{figure}

In this section we present a computational speed benchmark. We perform tests on the standard laptop with an Intel® Core™ i7-3537U processor and a GCC 6.3.1 C++ compiler. Both algorithms compute (\ref{eq:special_function_term}) for any $z$ by $10^4$ times. Our algorithm performs computations with accuracy $\epsilon = 10^{-10}$. Parameters are set to $\theta = 5.24$ and $\omega = 1.42$. GSL routines have a predefined accuracy and we can-not change it. Let us also mention that our algorithm can compute (\ref{eq:special_function_term}) for large values (e.g. $x > 732$) while GSL routines have overflow errors. Hence we need in some modifications of GSL routines (e.g. we can compute $e^{-x} \Psi(\theta, \omega,x)$ to avoid overflow) in case of large values of function argument. Figure (\ref{fig: speed_test}) illustrates a comparison between our method based on formulas (\ref{eq:series_exp_special_term} - \ref{eq:coef_c_and_d}) and direct computation of the numerator and denominator in (\ref{eq:special_function_term}) using GSL routines. Parameters are set to The left sub-figure illustrates the computational speed of algorithms based on our formula for small arguments (red line) and the GSL algorithm (blue line). For $0<x<1$ our algorithm is faster than GSL, but for $1<x<4$ GSL is faster. Let us note that if we change $\epsilon$ we will have other results. The right sub-figure illustrates speed's comparison in case of large argument $x$. In this case our solution is faster than GSL at whole segment $40<x<732$. For $732<x<1000$ GSL routines can not evaluate function value. We suggest the low speed of GSL routines may be caused by exponential grow of the Kummer functions for large arguments (it also can cause overflow errors). The source C++ codes can be found at GitHub repository (see link in the references).

\begin{figure}
	\begin{center}
		\resizebox*{18cm}{!}{\includegraphics{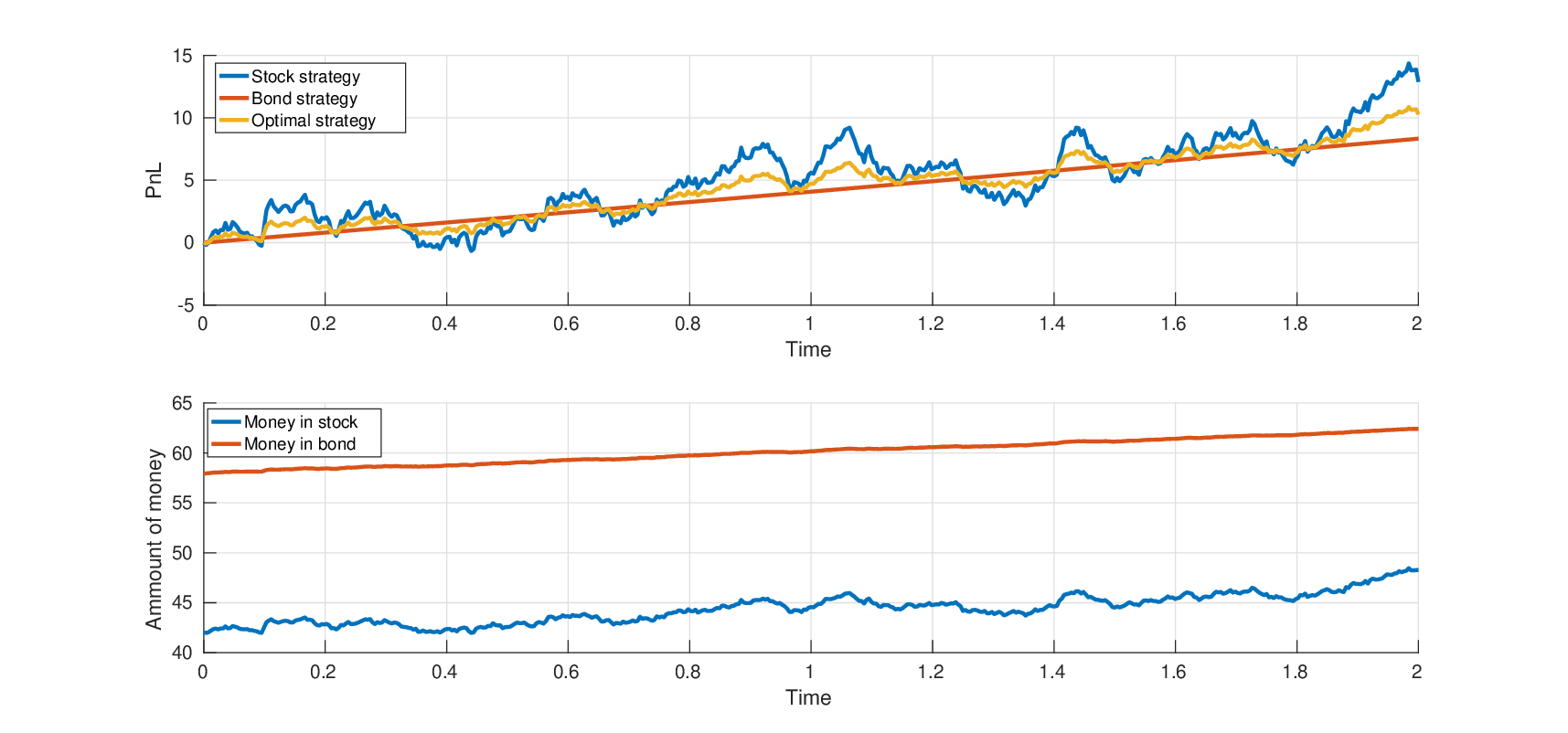}}
		\caption{\label{fig1} Comparison of different strategies. Positions in bond and stock in utility strategy.
			\label{fig: trajectory}}
	\end{center}
\end{figure}
\begin{figure}
	\begin{center}
		\resizebox*{18cm}{!}{\includegraphics{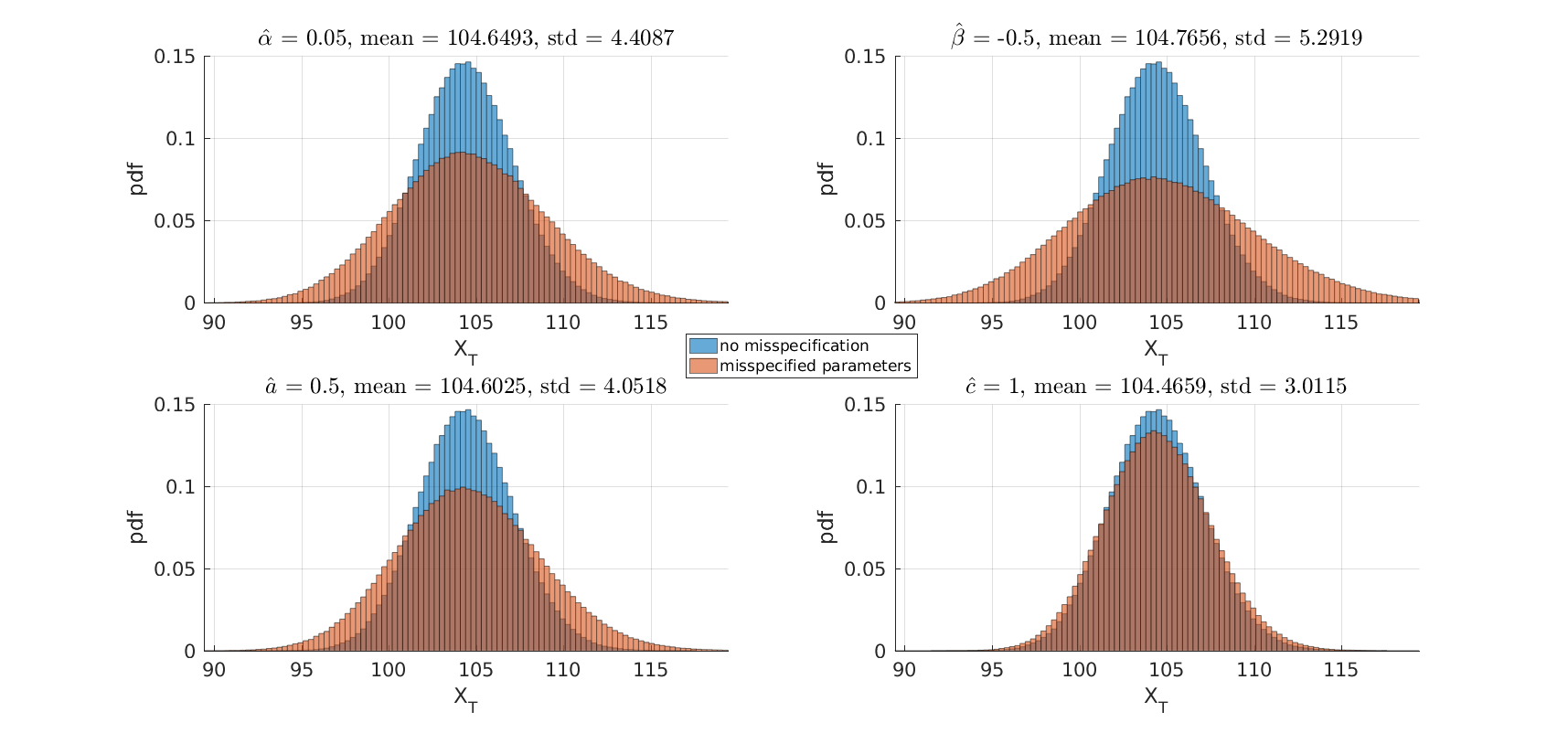}}
		\caption{\label{fig2} Misspecification of parameters. Blue distribution has $104.4291$ mean and $2.7365$ standard deviation.
			\label{fig: misspecification}}
	\end{center}
\end{figure}

\subsection{Parameters misspecification}
\par 
This section contains several numerical examples that illustrate optimal strategy and the effects of parameters' misspecification. Figure \ref{fig: trajectory}  demonstrates the wealth dynamics for 3 different investment strategies.The first strategy (PnL is colored by red) consists of only bond investments. We have invested all of the initial wealth $X_0 = 100$ in bond $B_t$ with initial value $B_0 = X_0$ and interest rate $r = 0.04$. The second strategy (blue line) consists of only stock investments. The stock process has initial value $S_0 = 100$, average return $\alpha = 0.045$, volatility $a = 0.4$, default intensity $c =0.8$ and skewness $\beta = -0.4$. In these strategies we do not have any portfolio re-balancing during the whole investing period $T=1$. Positions in the third strategy (yellow color) are defined by formulas \ref{eq:piStarExactMCEV}. The investor's risk aversion is $\gamma = -4$. The second figure \ref{fig: misspecification} illustrates the terminal wealth distribution with true and misspecified parameters. For these tests we have $10^6$ simulations to compute terminal wealth distributions. These examples show that calibration errors in average return $\alpha$ and skewness $\beta$ are more critical than errors in volatility level $a$ and default intensity $c$.   

\begin{figure}
	\begin{center}
		\resizebox*{18cm}{!}{\includegraphics{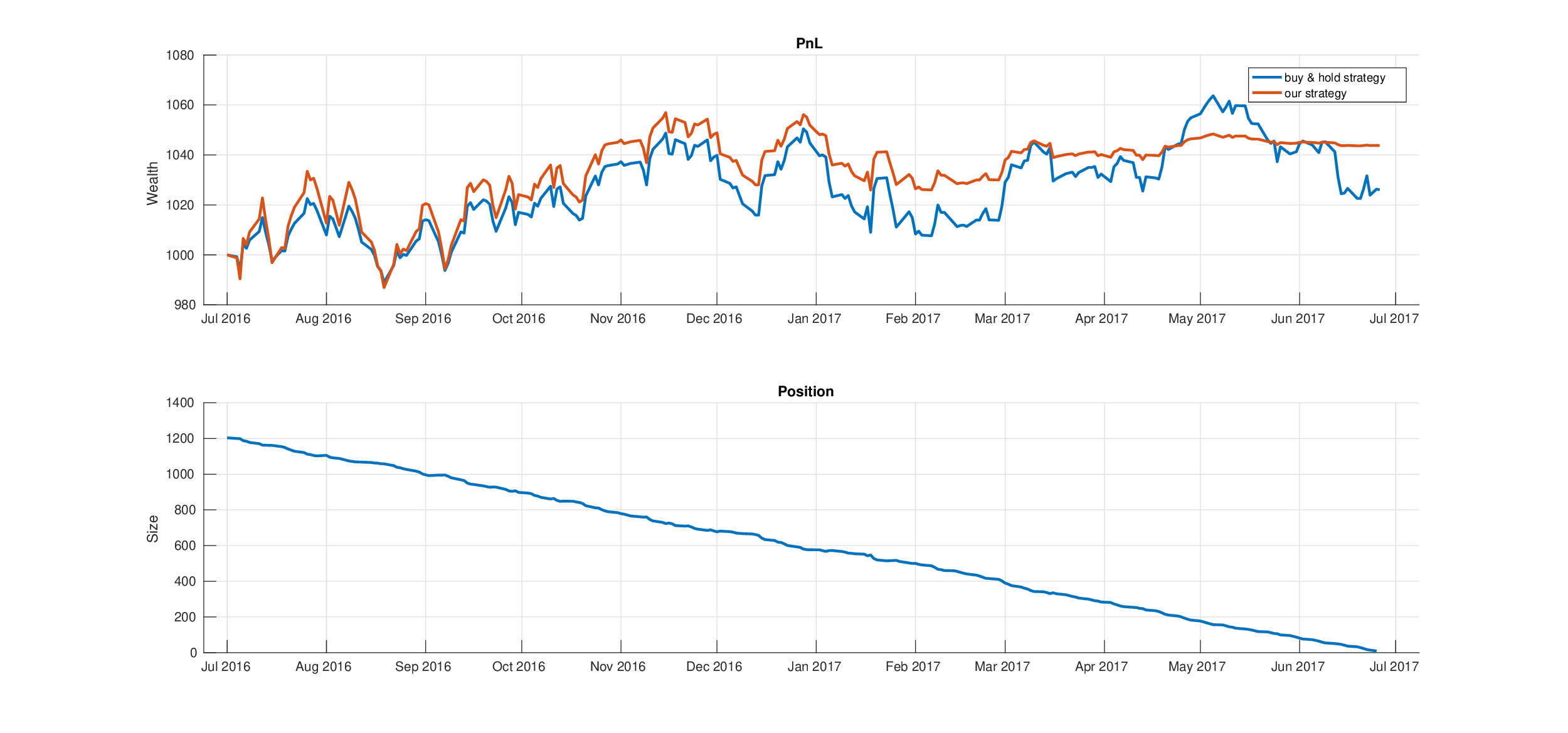}}
		\caption{\label{fig5} Comparison of different strategies on the USD/CAD FX Rates.      
			\label{fig: USDCAD strategy}}
	\end{center}
\end{figure}

\section{Applications to the algorithmic trading}\label{sec:pairtrading}
In this section we propose a statistical arbitrage strategy based on our obtained results. Consider an arbitrageur trading a mean-reverting asset. Suppose that the trader knows the 'fair' mean price of the asset (i.e. long term mean) and he knows that price will be return to this mean price. Generally, in this framework a trader can make profit by take a long position when the asset is below its long-term mean and a short when it is above. The question is in the size of the trader's position and how the position should be optimally managed depending on the price process parameters and trader's current wealth. This optimal trading problem can also be treated in the general portfolio optimization framework and it corresponds to the zero interest rates case i.e. we must set $r(t) = 0$ in all formulas. Therefore the wealth for a control $\pi$ is given by
\begin{eqnarray}
\label{eq:XdefTrading}
dX_s = \pi_s dS_s,
\end{eqnarray}
here $\pi_s$ is the trader position in the mean-reverting asset. 
It is well known that the original M-CEV price process $S_t$ can be mean-reverting if $\alpha < 0$. Without loss of generality, we consider the case of a square-root diffusion process which corresponds to parameters
\begin{eqnarray}
\label{eq:parametersforCIR}
\alpha = -\kappa, \quad\quad c = \frac{\kappa \bar{S}}{a^2}, \quad\quad \beta = -1/2.
\end{eqnarray}
This leads to the following mean reverting process
\begin{eqnarray}
\label{eq:CIRdef}
dS_s = \kappa (\bar{S}-S_s)dt + a\sqrt{S_s} dW_s, \quad S_t = S.
\end{eqnarray}
Parameter $\kappa$ is the reversion speed, $\bar{S}$ is long-term mean and $a$ is the volatility level. 
\begin{proposition}
For the process (\ref{eq:CIRdef}) the value function and optimal control allows representation (\ref{eq:exactfMCEV}) and (\ref{eq:piStarExactMCEV}). The parameters have the following representation
\begin{eqnarray}
\label{eq:parametersCIR}
\lambda =-\delta\frac{\kappa \bar{S}}{a^2}, \quad
\eta = \sqrt{\left(\lambda + \frac{1}{2}\right)^2  + \frac{ \delta (1-\delta) \kappa^2 \bar{S}^2}{a^8}}, \quad
R = 2\sqrt{\delta} \frac{\kappa \bar{S}} {a^2} \left( \delta + \frac{1-\delta}{a^2} \right)
\end{eqnarray}
Functions $A(\tau)$, $B(\tau)$ and $D(\tau)$ are given by
\begin{eqnarray}
\label{eq:ABDCIR}
A(\tau) = \frac{1}{2 \sinh^2(\tau) [\coth(\tau) + \sqrt{\delta}]}, \quad
B(\tau) = -\frac{1 - \delta}{2[\coth(\tau) + \sqrt{\delta}]}, \quad
D(\tau) = \sinh^2(\tau) [\coth(\tau) + \sqrt{\delta}]^2.
\end{eqnarray}
\end{proposition}
Optimal position is
\begin{eqnarray}
\label{eq:piStarExactCIR}
\pi^*(X,S,t) =  \frac{X}{S} \left( \delta  \frac{\kappa(\bar{S} -S) }{a^2}  + S B(\tau) + (\lambda + \eta + 1/2) \frac{M_{\lambda+1, \eta}\left( \frac{2\kappa \sqrt{\delta}}{a^2} S A(\tau)  \right)}{M_{\lambda, \eta}\left( \frac{2\kappa \sqrt{\delta}}{a^2} S A(\tau) \right)} \right).
\end{eqnarray}
If we intend to perform trading strategies based on optimal control of a square-root process we must properly construct the mean-reverting asset. One of the standard approaches to mean-reversion trading is pair trading. In this case we construct a mean-reverting asset as a difference(i.e. spread)  between two co-integrated assets. In almost all cases this spread has zero long term mean. Therefore we can not consider a square-root process for pair trading because it has only positive values if $2 \kappa \bar{S} > a^2$. This leads us to change the difference to another mean-reverting asset. We propose to make a strategy for FX rates. They are always positive and can be mean-reverting. Hence we can model it using a square-root process. Figure (\ref{fig: USDCAD strategy}) illustrates a trading strategy based on USD/CAD historical data. We consider daily data over 6.5-year time period from 01/01/2011 to 26/06/2017. We calibrate parameters of square-root process $\kappa$ ,$\bar{S}$ and $a$ on 01/01/2011-01/07/2016 daily rates. We obtain following values:     
\begin{eqnarray}
	\hat{\kappa} = 0.1090, \qquad \hat{\bar{S}} = 1.32675, \qquad a = 0.28789.
\end{eqnarray}
We test our strategy on the 01/07/2016-26/06/2017 time period, $T= 0.9961$. The investor's risk aversion is set to $\gamma = -7$, initial wealth is $X_0 = 1000$. Strategy based on formula (\ref{eq:piStarExactCIR}) has $4.33 \%$ return, $0.6464$ Sharpe ratio and $-6.54\%$ maximum drawdown,  while buy and hold (blue line) strategy has $2.61 \%$, $0.3911$ and $-7.03\%$.


\appendix
\section{Appendix: Proofs} \label{sec:appendix}
\subsection{Theorem \ref{thm:first}}
It is easy to show that unknown Green function $F_G(z,\tau; \xi)$ can be represented as
\[
F_G(z,\tau; \xi) = \left(z/\xi\right)^{\lambda} \exp\left\{R\tau + Q(z-\xi)/2\right\} h(z,\tau).
\]
where $z$ and $\tau$ are defined in (\ref{eq:zandtaudef}) and function $h(z,\tau)$ solves Cauchy problem ($\delta(z)$ is the Dirac delta function)
\begin{eqnarray}
h_{zz} + \left( -\frac{1}{4} + \frac{1/4 - \eta^2}{z^2} \right)h &=&\frac{1}{2z} h_{\tau},\\
h(z,0) &=& \delta(z-\xi) \left(\xi/z\right)^{\lambda} \exp\left\{ Q(\xi- z)/2\right\}.
\end{eqnarray}

Let $G(z;\zeta)$ be a Laplace transform of the function $h(z,\tau)$:
\[
G(z; \zeta) = \int_{0}^{\infty} e^{\zeta \tau} h(z, \tau) d \tau.
\]
It turns out to the following ODE for function $G$
\begin{eqnarray}
\label{eq:whittakereqdef}
G'' + \left(-\frac{1}{4}-\frac{\zeta/2}{z} + \frac{1/4-\eta^2}{z^2}\right)G = -\chi(z,\xi), \qquad  \chi(z,\xi) = \frac{1}{2z}\delta(z-\xi) \left(\xi/z\right)^{\lambda} \exp\left\{ Q(\xi- z)/2\right\}.
\end{eqnarray}
The homogeneous equation in (\ref{eq:whittakereqdef}) is called Whittaker equation and have two linearly independent solutions, namely $M_{-\zeta/2, \eta}(z)$ and $W_{-\zeta/2, \eta}(z)$ (see Abramowitz and Stegun (1973)).It is easy to show that the solution of non-homogeneous problem (\ref{eq:whittakereqdef}) can be represented as
\begin{equation}
\label{Eq: v_0}
G(z; \zeta) = \frac{1}{2\xi}\frac{\Gamma(1/2+\zeta/2+\eta)}{\Gamma(1+2\eta)} \left\{ {\begin{array}{l}
 M_{-\zeta/2,\eta}(z) W_{-\zeta/2,\eta}(\xi), \quad \xi \leq z\\
 M_{-\zeta/2,\eta}(\xi) W_{-\zeta/2,\eta}(z), \quad \xi \geq z\\	
 \end{array}} \right.
\end{equation}
Using this relation between Whittaker functions and modified Bessel function(see Gradshteyn and Ryzhik (1980), formula 6.669.4)
\begin{eqnarray}
\int_0^{\infty} e^{-\frac{1}{2} (a_1+a_2) t \cosh x} \coth ^{2\nu} \left( \frac{1}{2} x \right) I_{2\mu} \left(t \sqrt{a_1 a_2 } \sinh{x}\right) dx =
\frac{\Gamma \left( \frac{1}{2} +\mu - \nu \right)}{ t \sqrt{a_1 a_2} \Gamma(1+2\mu)} W_{\nu,\mu} (a_1 t) M_{\nu,\mu} (a_2 t), \nonumber\\
{\rm Re \,}  \left( \frac{1}{2} + \mu - \nu \right) > 0, \quad {\rm Re \,} \mu > 0, \quad a_1 >a_2.  \nonumber
\end{eqnarray}
we obtain new formula for $G(z,\zeta)$
\[
G(z;\zeta) = \frac{\sqrt{z/\xi}}{2} \int_{0}^{\infty} 
e^{- \frac{z + \xi}{2}\cosh{\psi}} \tanh ^{\zeta} \left( \frac{\psi}{2}\right)
I_{2\eta}\left( \sqrt{z \xi} \sinh{\psi}\right) d\psi.
\]
Next we introduce new integration variable $\nu$
\[
\log \left[ \tanh\left( \frac{\psi}{2}\right)\right] = \nu, \quad d\psi = \frac{d\nu}{\sinh(-\nu)}, \quad \psi = \frac{1}{\sinh\left( -\nu \right)}, \quad \cosh \psi = \coth \left( -\nu \right).
\]
In the result we have
\[
G(z; \zeta)=\frac{\sqrt{z/\xi}}{2} \int_{-\infty}^{0}
e^{-\frac{z+\xi}{2}\coth(-\nu) +\zeta \nu } I_{2\eta} \left( \frac{\sqrt{z\xi}} {\sinh(-\nu)}\right)  \frac{d\nu}{\sinh(-\nu)}.
\]
Inverting the Laplace transform, we recover the formula for $h(z,t)$
\begin{eqnarray}
\label{eq:fLaplaceInversion}
h(z,t) = \frac{\sqrt{z/\xi}}{4\pi i} \int_{N-i\infty}^{N+i\infty} \int_{-\infty}^{0}
e^{-\frac{z+\xi}{2}\coth(-\nu) +\zeta (\nu + \tau) } I_{2\eta} \left( \frac{\sqrt{z\xi}} {\sinh(-\nu)}\right)  \frac{d\zeta d\nu}{\sinh(-\nu)}
\end{eqnarray}
where $N$ is a number such that all residues of the integrand are to the right of it.
Using the well-known representation of Dirac function
\[
\frac{1}{2\pi i} \int_{N-i\infty}^{N+i\infty} e^{z \zeta} d\zeta = \delta(z),
\]
and changing the order of integration in (\ref{eq:fLaplaceInversion}), we get
\begin{eqnarray}
\label{eq:fLaplaceInversion1}
h(z,t) = \frac{\sqrt{z/\xi}}{2} \int_{-\infty}^{0}
\delta(\nu + \tau) e^{-\frac{z+\xi}{2}\coth(-\nu)} I_{2\eta} \left( \frac{\sqrt{z\xi}} {\sinh(-\nu)}\right)  \frac{d\nu}{\sinh(-\nu)}
\end{eqnarray}
Note, that $\tau \geq 0$. Thus, we can complement the range of integration in (\ref{eq:fLaplaceInversion1}) to the whole line, and, using the definition of Dirac's function, namely  $\int_{-\infty}^{\infty} \delta(\zeta -z) u(\zeta)d\zeta = u(z)$ for any continuous $u$, we get the main formula (\ref{eq:Green_function_main_formula}) for $F_G$.
\subsection{Theorem \ref{thm:second}}
Consider the quotinent of two series
\[
\sum_{s=0}^{\infty} c_s x^s= \frac{\sum_{s=0}^{\infty} a_s x^s}{\sum_{s=0}^{\infty} b_s x^s}
\] 
It is equivalent to 
\[
\sum_{s=0}^{\infty} c_s x^s  \sum_{s=0}^{\infty} b_s x^s = \sum_{s=0}^{\infty} a_s x^s\] 
or 
\[
\sum_{s=0}^{\infty} \left( \sum_{i+j = s} b_i c_j x^{i+j}\right) = \sum_{s=0}^{\infty} \left( \sum_{i+j = s} b_i c_j x^{s}\right)=  \sum_{s=0}^{\infty} a_s x^s\] 
Hence the coefficients $c_s$ solve the following linear system:
\begin{eqnarray}
c_0 b_0 &=& a_0, \nonumber \\
c_0 b_1 + c_1 b_0 &=& a_1, \nonumber \\
c_0 b_2 + c_1 b_1 + c_2 b_0 &=& a_2, \nonumber \\ 
&...& \nonumber \\
\sum_{i+j = k} c_i b_j &=& a_k, \nonumber \\ 
&...&
\end{eqnarray}
For $c_s$ in \ref{eq:coef_c_and_d} we use the following definition of Kummer function (see Abramovitz and Stegun (1972))
\begin{eqnarray}
\Psi(\theta, \omega, x) = \sum_{s=0}^{\infty} \frac{(\theta)_s}{(\omega)_s s!} x^{s} = 1+ \frac{\theta}{\omega} x + \frac{\theta(\theta + 1)}{\omega(\omega + 1) 2!} x^2 + ...
\end{eqnarray}
and for $d_s$ we use the asymptotic of Kummer function for large argument (see Abramovitz and Stegun (1972))
\begin{eqnarray}
\label{eq:Kummer_Large_x}
\Psi(\theta, \omega, x) \thicksim \frac{e^x x^{\theta - \omega}}{\Gamma(\theta)}\sum_{s=1}^{\infty} \frac{(1- \theta)_s (\omega - \theta)_s}{s!} x^{-s}, \quad\quad x\rightarrow\infty.
\end{eqnarray}
\end{document}